\documentstyle [11pt,amsbsy] {article}
\date {} 			 
\renewcommand {\vec} {\boldsymbol}
\begin {document} 
\everymath {\displaystyle}

\medskip
\centerline {\Large \bf About spin precession in electromagnetic
wave} \medskip \centerline { \large Sergey L. Cherkas} \medskip
\centerline {{\it Institute of Nuclear Probems,
Minsk 220050, Republic of Belarus} }
\bigskip
\centerline {\bf Abstract}
\smallskip
\par
\noindent
{ \small
It is shown that two ways of the
description of the $ {1/2} $--spin precession in a 
circularly polarized electromagnetic wave: on the basis of the 
Bargmann-Michel-Telegdi equation and on the basis of the Dirac 
equation with Pauli term give the same result in the second order on 
the wave's field for arbitrary wave length.  } \medskip \bigskip

\par \noindent
{ \bf 1. Introduction.}
The spin rotation of a particle in an electromagnetic wave is
a perspective method of research of the electromagnetic and
weak interactions.
In recent work [1]
electron spin rotation
was considered, when electron moves towards
circularly polarized electromagnetic wave. It was shown, that
because of  quantum electrodynamic effects frequency of spin rotation
is much more than it would follow from the
Bargmann-Michel-Telegdi (BMT) equation. The account
quantum electrodynamic  effects means, that an electron
is considered
as not a dot
particle but as a particle having
some electromagnetic "structure" due to absorption and
radiation of the virtual $--\gamma $ quanta.
Similar effects were
discussed in [2] where
the polarization rotation of
 $ \gamma $--quanta in the medium with
polarized electrons is considered.
The condition, at whichn  the  effect becomes
essential, looks like [1] $ {4\varepsilon \omega \over m ^ {2}} \ge
1 ~, $
where $ \omega $ is a frequency of the electromagnetic wave,
$m $ is electron mass ( system of units $ \hbar=c=1 $ is used) .
Under this condition
energy of the photon emitted becomes comparable
with the electron  energy $ \varepsilon $,
however, under
the same condition
quasi-classical description of the moving particle becomes
inapplicable  and the concept
  of a trajectory loses sense.
The description used in [1] kept features
of the quasi-classical
 description, typical for the
BMT equation.
If to look at the Dirac equation for a free
electron, we find out, that free electron is not a dot
particle too i.e. it has some "structure".   Let's following [3]
consider
wave packet with the average momentum equal to
zero. If to require, that the wave packet is a superposition
of the solutions only with positive energy, we shall notice, that
we cannot localize it in the area about $1/m $ size. To
locate electron in the range of the order $1/m $ it is
necessary to add to wave function solutions with negative
energy. Electron localized such a way becomes
"electron-positron".  Concept of a trajectory (and the
BMT equation) cannot be used if in the electron rest system (where
electron average momentum is equal to zero)
localization  length
$1/m $ greater than the inverse
electromagnetic wave frequency
$1/\tilde \omega $ in the
electron rest system.
Passing to the laboratory system we shall receive
for the relativistic
electrons the inequality above within a factor of $2$.
That is Dirac structure of the electron becomes
essential under the same condition, that electromagnetic one.
The question arises about Dirac structure influence on
the frequency of the spin rotation compared with the BMT equation.
To learn it, we shall calculate frequency of rotation of the
electron spin in a circularly polarized wave,
proceeding from the BMT equation and from the
Dirac equation with the anomalous magnetic moment. Running
forward we should tell, that both methods give identical result in
the second order on the wave field strength, i.e.  BMT equation
casually gives correct result even behind the area
of the  quasi-classical description applicability.

\bigskip
{ \bf 2. The description of a spin rotation in a frame of
BMT equation.}

Let us present field of the circularly-polarized wave
propagating
against axis $z $ as:
\begin {eqnarray}
H_x=H_0\cos\xi ~, ~~~~~~~ H_y=H_0\sin\xi
~, ~~~~~~~\vec E =\vec e_z\times \vec H
\nonumber \\
\xi\equiv k_\mu x ^\mu\equiv \omega (t+z)
~, ~~~~~~~ A_\mu=a_\mu e ^ {-i\xi} +a ^ *_\mu e ^ {i\xi} ~.
\label {2}
\end {eqnarray}
Here $ \vec H, \vec E $ are srength of the magnetic and
electrical fields,
$A_\mu $ is 4-potential of the circularly polarized wave, $
\omega $ is wave frequency. The constant 4-vector $a_\mu $
is:  $a_\mu = (0, a, ia, 0) $, where $a =-H_0/2\omega $.
Consider electron motion in the circularly polarized
wave.  The solution, for which momentum component
perpendicular to $z $ axis is equal to zero in
average, looks like [4]:
\begin {eqnarray} \vec p_\bot = -e\vec A
(\xi) ~, ~~~~~ p_z=const ~, ~~~~~
\varepsilon^2=p_z^2+e^2A^2+m^2=const, \nonumber \\ z = \frac {p_z t}
{\varepsilon} ~, ~~~ \vec v_\bot =\frac {\vec p_\bot} {\varepsilon} =
\frac {e\vec H (\xi)} {\varepsilon\omega} ~, ~~~~\vec
r_\bot=-\frac{e}{\omega(\varepsilon+p_z)} \int \vec A (\xi) d\xi ~.
\end {eqnarray}
Substituting speed $ \vec v $ to the BMT equation
[5]
\begin {equation} { d {\vec \zeta} \over
dt} = {2\mu m+2\mu ' (\varepsilon -m) \over \varepsilon} ({\vec
\zeta} \times \vec H) + {2\mu '\over \varepsilon +m} (\vec v\cdot
\vec H) (\vec v\times {\vec \zeta}) + {2\mu m+2\mu '\varepsilon \over
\varepsilon +m} {\vec \zeta} \times (\vec E\times \vec v) ~, \label
{4}
\end {equation}
we receive in the second order on the field strength:
\begin {eqnarray}
( \vec v\cdot \vec H) (\vec v\times \vec \zeta) = (\vec v_\bot\cdot
\vec H) (
\vec v_\parallel +\vec v_\bot) \times \vec\zeta
 \approx {{ev_\parallel} \over {\varepsilon
\omega}} H ^ {2} (\vec e _ {z} \times \vec {\zeta}) ~, ~~~~~~~
\nonumber
\\
\vec {\zeta} \times (\vec E\times
\vec v) = \vec {\zeta} \times
( \vec E\times \vec v_\parallel) + \vec {\zeta} \times (\vec E\times
\vec v_\bot) =v_\parallel
( \vec {\zeta} \times \vec H) - (\vec {\zeta} \times
\vec e _ {z}) H ^ {2} {e ^ {2} \over {\varepsilon \omega}}.
\label {5}
\end {eqnarray}
Having substituted (\ref {5}) in (\ref {4}) we have:
\begin {eqnarray}
\frac {d\vec {\zeta}} {dt}
 &=&\Biggl (2\mu ^\prime \left (1 +
\frac {\sqrt{\varepsilon^2+m^2}}{\varepsilon}\right)+
\frac {e} {\varepsilon}
\left (
 1+\sqrt{\frac{\varepsilon-m}{\varepsilon+m}}\right)\Biggr)
( \vec \zeta\times
\vec H) \nonumber \\
 &&~~~~~
-\Biggl (\frac {2\mu ^\prime e} {\varepsilon\omega}
 \left(1+\sqrt{\frac{\varepsilon-m}{\varepsilon+m}}\right)
+ \frac {e^2} {\varepsilon (\varepsilon +m) \omega} \Biggr)
H^2 (\vec \zeta\times \vec e_z)
\label {6}
\end {eqnarray}
The equation (\ref {6}) contains at the right hand side
sum
of the oscillating and constant terms, however one
cannot
reject oscillating term simply.
Let's
share the vector $ \vec {\zeta} $ into two parts [6]:
\begin
{equation} \vec {\zeta} (t) = \vec {\theta} (t) + \vec {\ae} (t),
\end {equation} where $ < \vec {\ae} (t) > =0 $
(angular
brackets mean averaging on interval of time much more
exceeding the period of the electromagnetic wave).
For $ \vec
{\theta} (t) $ and $ \vec {\ae} (t) $
we may write
\begin
{eqnarray} { d\vec {\theta} \over dt}&=& \Biggl (2\mu ^\prime \left
(1 + \frac {\sqrt{\varepsilon^2+m^2}}{\varepsilon}\right)+ \frac {e}
{\varepsilon} \left (
 1+\sqrt{\frac{\varepsilon-m}{\varepsilon+m}}\right)\Biggr)
< \vec \ae\times \vec H >
\nonumber \\&&~~~~~~~~~
-\Biggl (\frac {2\mu ^\prime
e} {\varepsilon\omega}
 \left(1+\sqrt{\frac{\varepsilon-m}{\varepsilon+m}}\right)
+ \frac {e^2} {\varepsilon (\varepsilon +m) \omega} \Biggr)
H^2 (\vec \theta\times \vec e_z)
\label {8}
\\
{ d\vec {\ae} \over dt}&=&
\Biggl (2\mu ^\prime \left (1 +
\frac {\sqrt{\varepsilon^2+m^2}}{\varepsilon}\right)+
\frac {e} {\varepsilon}
\left (
 1+\sqrt{\frac{\varepsilon-m}{\varepsilon+m}}\right)
\Biggr) (\vec \theta\times
\vec H)
\label {9}
\end {eqnarray}
In the second equation
the terms
proportional to
$ \left (\vec {\ae} \times \vec H- <\vec {\ae} \times \vec H > \right) $
   and
$H ^ {2} (\vec {\ae} \times \vec e _ {z}) $ are rejected,
 because they give
contribution higher than second order on the field strength to
 the equation for $ {d\vec {\theta} \over dt} $.  Integrating the
equation (\ref {9}) we get
\begin {eqnarray} \vec {\ae}
(t) = \frac {H _ {0} \varepsilon} {2\omega (\varepsilon +p _ {z})}
\Biggl (2\mu ^\prime \left (1 + \frac
{\sqrt{\varepsilon^2+m^2}}{\varepsilon}\right)+ \frac {e}
{\varepsilon} \left (
 1+\sqrt{\frac{\varepsilon-m}{\varepsilon+m}}\right)\Biggr)
\vec {\theta} \times (\vec e _ {x} \sin \xi
-\vec e _ {z} \cos\xi).
\label {10}
\end {eqnarray}
Having substituted (\ref {10})
in (\ref {8})
we arrive at
\begin {eqnarray}
 {d\vec {\theta} \over dt} =
2\mu ^ {\prime 2} ~ \frac {\varepsilon + \sqrt {\varepsilon^2+m^2}}
{ \varepsilon \omega} H^2 (\vec \theta\times \vec e_z) ~,
\nonumber
\\
{ d\vec {\theta} \over dz} =
\frac {\varepsilon} {p_z}
\frac {d\vec \theta} {dt} =
\frac {2\mu ^ {\prime 2}} {\omega} ~
 \frac{\varepsilon+\sqrt{\varepsilon^2+m^2}}
{ \sqrt {\varepsilon^2+m^2}} H^2 (
\vec
\theta\times \vec e_z).
\label {11}
\end {eqnarray}
It is necessary to note, that generalization of the BMT equation
on a case of non-uniform fields is deduced in [7,8].
The equation [7,8] is mysterious, in the sense, that it fails to be
deduced from the Dirac equation with the anomalous
magnetic moment,
using Foldi-Woithausen transformation
[9,10], however it can be deduced from the Dirac
equation applying the correspondence principle [11,12,13].  Frequency
of spin rotation received from the equation [7,8] contains in
comparison with (\ref {11}) additional terms.

\bigskip
{ \bf 3. Description of the spin rotation in electromagnetic
wave, proceeding from the Dirac equation.  }
Let us proceed from the assumption, that in a field of
the electromagnetic
wave electron has some spin-dependent effective energy, which
results to different  refractive indexes for electrons with
the different spin projections to the momentum direction.
To describe
it we receive the equation for the
electron wave function
averaged over phase of
the electromagnetic
wave. Averaged electron wave function
is a plain wave with a wave 4-vector satisfying to the
dispersion equation.
 \par Write down equation for
bispinor $ \Psi (x, b) $ of the electron  in the electromagnetic wave
as:
\begin {equation} \alpha _ {x-b} \Psi (x, b) = 0, \label {12}
\end {equation}
Where $x = \{t, \vec r \} $, and $b $ is a constant 4-vector
connected with the phase of the
electromagnetic wave $ \phi _ {0}$ as $\phi _ {0}
=k ^\mu b_\mu $.
Averaging of wave function $ < \Psi (x) > = \frac
{1} {\Omega} \int_\Omega\Psi ( x, b) d ^ {4} b $ is equivalent to
the averaging over phase of the wave.
 Here $ \Omega $ is some
large 4-volume. Average wave function can be presented as:
\begin{equation} < \Psi (x) > =e^{-{ip'x'}} u (p ')~, \label{14}
\end{equation} here $u (p ') $ is bispinor, and $p ^\prime
$ is 4-quasi-momentum required
for which we are going to
deduce dispersion equation.  Let's designate $
{\overline\alpha} (p ^\prime) \equiv \frac {1} {\Omega} \int_\Omega e
^ {-ip ^\prime x} \alpha_x e ^ {ip ^\prime x} dx^4 $
and rewrite
(\ref {12}) as:
\begin {equation} {\overline\alpha} {\Psi (x, b)} =
({\overline\alpha} - \alpha _ {x-b}) \Psi (x, b) = {0.} \label {16}
\end {equation}
The averaging of (\ref {16}) on $b $ gives
\begin
{eqnarray} {\overline\alpha} < \Psi (x) > = \frac {1} {\Omega}
\int_\Omega ({\overline \alpha} -
 \alpha_{x-b})\Psi(x,b)d^4b=\frac{1}{\Omega}\int_\Omega
( {\overline \alpha} -\alpha _ {x-b}) e ^ {-ip ^\prime b} \Psi (x-b, 0) d^4b
\nonumber \\ =
\frac {\exp (-ip ^\prime x)} {\Omega} \int_\Omega e ^ {ip ^\prime x ^\prime}
( {\overline \alpha}-\alpha_{x^\prime})\Psi(x^\prime,0)d^4x^\prime~~.
\label {17}
\end {eqnarray}
Under derivation of (\ref {17}) we take into account that by virtue
of space - time uniformity $ \Psi (x, b) $ has the following
transformation properties [14]:  $ \Psi (x+b ^\prime, b+b
^\prime) =e ^ {-ip ^\prime b ^\prime} \Psi (x, b) $, where $b' $ is
any 4-vector.  Presenting $ \Psi (x, 0) $ as $ \Psi (x, 0) = <
\Psi (x) > + \phi (x) $ and substituting it into the (\ref {17}) we
find \begin {equation} \bar \alpha < \Psi (x) > = \exp (-ipx) Z (p
^\prime) ~, \label {19} \end {equation} where $ Z (p ^\prime) = \frac
{1} {\Omega} \int_\Omega e ^ {ip ^\prime x ^\prime} ( \bar \alpha
-\alpha _ {x ^\prime}) \phi (x ^\prime) dx ^ {\prime 4} ~.  $ Writing
down (\ref {12}) For $b=0 $ and subtracting it from (\ref {19}) we
obtain \begin {equation} ( \bar \alpha - \alpha_x) < \Psi (x) >
-\alpha_x\phi (x) = e ^ {-ip ^\prime x} Z (p ^\prime) ~.  \label {20}
\end {equation}
Considering $Z (p ^\prime) $ as known quantity we
solve (\ref {20}) for $ \phi (x) $, and substituting $
\phi (x) $ again to the definition of $Z (p ^\prime) $ find $Z
(p ^\prime) $
\begin {eqnarray} Z(p^\prime)&=&\left(1 -\frac {1}
{\Omega} \int_\Omega e ^ {ip ^\prime x ^\prime} ( \bar \alpha -\alpha
_ {x ^\prime}) { 1\over \alpha _ {x ^\prime}} e ^ {-ip ^\prime x
^\prime} d^4x ^\prime \right) ^ {-1} ~~~~~~~~ \nonumber \\ &&
~~~~~~~\times
\frac {1} \Omega \int_\Omega e ^ {ip ^\prime x}
( \bar \alpha -\alpha_x) {1\over \alpha_x}
( \bar \alpha -\alpha_x) < \Psi (x) > d^4x ~.
\label {22}
\end {eqnarray}
Substituting $ < \Psi (x) > $  in the form (\ref {14})
to (\ref {22})
and  then
substituting (\ref {22}) in (\ref {19})
the dispersion
equation
is received
\begin{eqnarray} \lefteqn{ \Biggl(\bar \alpha(p^\prime)-
\left(1 -\frac{1}{ \Omega }
\int_\Omega e^{ip^\prime x^\prime}
(
\bar \alpha -\alpha _{x^\prime})
{1\over \alpha_{x^\prime} }e^{-ip^\prime x^\prime}d^4x^\prime
\right)^{-1}~~~~~~~~
}
\nonumber
\\ & &
~~~~~~~\times
\frac{1}\Omega \int_\Omega e^{ip^\prime x}
(\bar \alpha -\alpha_x){1\over \alpha_x}
(\bar \alpha -\alpha_x)e^{-ip^\prime x}
d^4x\Biggr)u(p^\prime)~=~0~.
\label{23}
\end{eqnarray}
For the Dirac equation with Pauli
anomalous magnetic moment $ \mu
^\prime $
operator $ \alpha _ {x} $ is
\begin {equation} { \alpha _ {x}} =
\gamma_\mu ( p ^\mu-eA ^\mu (x)) -m +\frac {i} {2} \mu ^\prime\sigma
_ {\mu\nu} F ^ {\mu\nu} (x) ~, \end {equation} Where $F^{\mu\nu} (x)
= \partial ^\mu A ^\nu-\partial ^\nu A ^\nu ~ $; $ \sigma _ {\mu\nu}
= {1\over 2} (\gamma _ \mu \gamma _ \nu -\gamma _ \nu \gamma _ \mu) ~
$; $p ^\mu = \{i {\partial \over \partial t}, -i\vec \nabla \} $.

In the second order on the field strength dispersion equation will be
\begin{eqnarray}
\lefteqn{
\Biggl(
\gamma^\mu p^\prime_\mu-m
-\frac{1}{\Omega }
\int_\Omega e^{ip^\prime x}\left(
e\gamma_\mu A^\mu (x)
-\frac{i}{2}\mu^\prime\sigma_{\mu\nu}
F^{\mu\nu}(x)\right)
G_0(x,x^\prime)~~~~~
}
\nonumber
\\ & &
~~~~~~\times
\left(e\gamma_\mu A^\mu (x^\prime)
-\frac{i}{2}\mu^\prime\sigma_{\mu\nu}
F^{\mu\nu}(x^\prime)\right)
e^{-ip^\prime x^\prime}d^4xd^4x^\prime
\Biggr)u(p^\prime)~=~0~.
\end{eqnarray}
Green function [5] of the free
Dirac equation $G _ {0} (x, x ') $ is determined as
\par
\noindent
$G _ {0}
( x, x ^\prime) = \int {G (p) \over
(2\pi) ^4} e ^ {-ip (x-x ^\prime)} d ^ {4} p $,
$ G (p) = \frac {\gamma_\mu p ^\mu
+m} {p^2-m^2} $.
Taking 4-potential $A ^\mu $ in the form (\ref {2})
we find
\begin {eqnarray} e\gamma_\mu A ^\mu
(x) -\frac {i} {2} \mu ^\prime\sigma _ {\mu\nu} F ^ {\mu\nu} (x) =
\left (e\gamma_\mu a ^\mu -\frac {\mu ^\prime} {2} \sigma _ {\mu\nu} (k ^\mu
a ^\nu -k ^\nu a ^\mu) \right) e ^ {-i\xi} + \nonumber \\ \left (e\gamma_\mu
a ^ {*\mu} + \frac {\mu ^\prime} {2} \sigma _ {\mu\nu} (k ^\mu a ^ {*\nu} -k ^\nu
a ^ {*\mu}) \right) e ^ {i\xi} \equiv W ^ {-} e ^ {-i\xi} +W ^ {+} e ^ {i\xi}
\label {27}
\end {eqnarray}
Let's consider integral:
\begin {eqnarray}
\frac {1} {\Omega} \int_\Omega e ^ {ip ^\prime x} \left (
 W^{-}e^{-ikx}+W^{+}e^{ikx}\right)G(p)e^{-ip(x-x^\prime)}
 \left(W^{-}e^{-ikx^\prime}+W{+}e^{ikx^\prime} \right)
e ^ {ip ^\prime x} d^4xd^4x ^\prime\frac {d^4p} {(2\pi) ^4}
\nonumber
\\
=
\frac {(2\pi) ^4} {\Omega} \int \left (W ^ {-}\delta^4 (p+k-p ^\prime) +
W ^ {+}\delta (p-k-p ^\prime) \right)
G (p)
\nonumber
\\ \times
 \left(W^{-}\delta^4(p-k-p^\prime)+W^{+}\delta^4(p+k-p^\prime)
\right) d^4p
W ^ {-} G (p ^\prime-k) W ^ {+} +
W ^ {+} G (p ^\prime+k) W ^ {-} ~.
\label {28}
\end {eqnarray}
At derivation of (\ref {28}) we use, that
$
 \left(\delta^4(p)\right)^2=\frac{1}{(2\pi)^4}\int
 e^{ipx}d^4x\delta^4(p)=\frac{\Omega}{(2\pi)^4}\delta^4(p)~.
$
This gives
following matrix equation in the second order on a field
strength
\begin {equation}
\left (\gamma ^\mu p ^\prime_\mu -m-
 W^{-}G(p-k)W^{+}+W^{+}G(p+k)W^{-}\right)u(p^\prime)=0
\label {29}
\end {equation}
The determinant of the equation (\ref {29}),
if to take standard
matrixes representation [5] is:
\begin {equation}
\det\left\vert \,\matrix {
 \varepsilon^{-}-A_1(\omega)&0&-p^\prime-B(\omega)&0 \cr
 0&\varepsilon^{-}-A_1(-\omega)&0&p^\prime+B(-\omega)\cr
 p^\prime+B(\omega)&0&-\varepsilon^{+}-A_2(\omega)&0 \cr
 0&-p^\prime-B(-\omega)&0&-\varepsilon^{+}-A_2(-\omega)\cr
} \right\vert ~~~~.
\label {30}
\end {equation}
Here we have designated: $ \varepsilon ^\pm =\varepsilon\pm m $,
\begin {eqnarray}
 A_1(\omega)&=&-\frac{2a^2}{\omega(\varepsilon+p)}
 \left(e^2(\varepsilon-\omega-m)-2e\mu^\prime
\omega (\varepsilon-m+p) +2\mu ^ {\prime 2} \omega^2
(\varepsilon+p) \right) ~,
\nonumber
\\
 A_2(\omega)&=&-\frac{2a^2}{\omega(\varepsilon+p)}
 \left(e^2(-\varepsilon+\omega-m)-2e\mu^\prime
\omega (\varepsilon+m+p) -2\mu ^ {\prime 2} \omega^2
(\varepsilon+p) \right) ~,
\nonumber
\\
 B(\omega)&=&-\frac{2a^2}{\omega(\varepsilon+p)}
\left (-e^2 (p +\omega) +2e\mu ^\prime
\omega m+2\mu ^ {\prime 2} \omega^2
(\varepsilon+p) \right) ~.
\nonumber
\end {eqnarray}
Evaluating the determinant (\ref {30}) and equating it to zero
the dispersion equation is obtained
\begin{eqnarray} \lefteqn{
\left( (\varepsilon^{-}-A_1(\omega))
(\varepsilon^{+}+A_2(\omega))-(p^\prime+B(\omega))^2\right)
}
\nonumber
\\ & & \times
\left( (\varepsilon^{-}-A_1(-\omega))
(\varepsilon^{+}+A_2(-\omega))-(p^\prime+B(-\omega))^2\right)
=0~.
\end{eqnarray}
For the solutions with positive $p ^\prime $ (we remind, that
electron goes in a positive direction of $z$ axis ) we have:
\begin {eqnarray}
 p_1^\prime&=&-B(\omega)+\sqrt{(\varepsilon^{-}-A_1(\omega))
( \varepsilon ^ {+} + A_2 (\omega))}
\nonumber
\\
 p_2^\prime&=&-B(-\omega)+\sqrt{(\varepsilon^{-}-A_1(-\omega))
( \varepsilon ^ {+} + A_2 (-\omega))} ~.
\nonumber
\end {eqnarray}
Two different helicities
correspond to  the two different electron
wave 3-vectors.
By taking their difference [15] we receive
frequency of spin rotation $ {d\theta \over dz} $, which in the
second order on a field strength coincides with received from the
formula (\ref {11}).

\bigskip
{ \bf 4. How quantum electrodynamics effects can be taken into
account.  } First of all we note, that there is a way to find photon
medium refractive index which take into account QED effects, with the
help of photon-electron forward scattering
amplitude
[16].  On physical sense
this way corresponds to the situation, when electron moves through
non-coherent photon medium, while
in our case photons are in a coherent state.
Let's show
how to receive dispersion equation in our approach with the
account of QED effects.
Under deduction of the dispersion equation
we proceeded from the equation:
\begin {equation} \alpha _ {x} \Psi
(x) =0.
\end {equation}
The equation it is possible to rewrite as:
\begin {equation} G ^ {-1} _ {x} \Psi (x) =0, \label {227}
\end {equation}
where $G _ {x} $ is a Green  function of the operator
$ \alpha _ {x} $.  To take into account QED effects, it is
necessary, keeping form of the equation (\ref {227}) to take as
$G _ {x} $ one-partial
QED Green
function of the electron
in the field of the electromagnetic wave.
As it is known [5],
$G _ {x} $ kernel looks like
\[ G (x, x ^\prime) = -i \frac {<
0|T\psi (x) \bar \psi (x ^\prime) S|0 >} {< 0|S|0 >} ~, \]
where
$
S=Texp\left (-ie\int\bar\psi (x) A ^\mu (x) \gamma_\mu\psi (x)
d^4x\right) ~, $ $ \psi (x) = \sum_n
a_n\psi_n^{(+)}(x)+b_n^{+}\psi_n^{(-)}(x)~, $ \par \noindent $ \bar
\psi (x) = \sum_n a_n^{+}\bar\psi_n^{(+)}(x)+b_n\bar\psi_n^{(-)}(x)~.
$
Here $A ^\mu (x) $ is an operator
 of the electromagnetic field quantized.
The external field is taken into account in the operators
$ \psi (x) $, $ \bar\psi (x) $, since
$ \psi ^ {(+)} _ {n} (x) $ and $ \psi
 ^ {({-})} _ {n} (x) $ is exact electron and
positron solutions of the Dirac equation in a field of an
electromagnetic wave ( the Volkov solution [5]).
Substituting $G ^ {-1} _ {x} $ in (\ref {23}) instead of the operator
$ \alpha _ {x} $ we find:
\begin {eqnarray} \Biggl (\gamma_\mu
p ^\prime_\mu -m - {\overline M} - \left (1-\frac {1} {\Omega}
\int_\Omega e ^ {ip ^\prime x ^\prime} ( M _ {x ^\prime} - {\overline
M} +eA_\mu\gamma ^\mu) G _ {x ^\prime} e ^ {-ip ^\prime x ^\prime}
d^4x ^\prime \right) ^ {-1} \nonumber \\ \times\frac {1} {\Omega}
 \int_\Omega e ^ {ip ^\prime x} (M_x - {\overline M} + eA_\mu\gamma
^\mu) G_x (M_x- {\overline M} +eA_\mu\gamma ^\mu) e ^ {-ip ^\prime x}
d^4x\Biggr) u (p ^\prime) =0 ~.  \label {288} \end {eqnarray}
Here we
have designated $M _ {x} = \pmatrix {G ^ {e} _ {x}} ^ {-1} - G ^ {-1}
_ {x} $, where $G ^ {e} _ {x} $ Green function of the Dirac
equation (without the anomalous magnetic moment) in a field
of a wave and \par $ { \overline M} (p ^\prime) \equiv {1\over
\Omega} \int_\Omega e ^ {-ip ^\prime x} M_xe ^ {ip ^\prime x} d^4x ~.
$
To carry out calculations with the help of
the dispersion
equation (\ref {288}) it is possible
to take the mass operator
$M _ {x} $ of a particle in field of an electromagnetic wave
calculated in works [17,18,19].

\bigskip
{ \bf 5. The conclusion.}
So, we have shown that
two ways of calculation of the spin rotation frequency of a
particle driven towards circularly
polarized electromagnetic wave :  Proceeding from the BMT equation
and from the Dirac equation give identical results in the
second order on a field strength.  Let's notice, that the
equation BMT turns out from the Dirac equation at the
following assumptions:  \par a) From Dirac hamiltonian
acting in basis of electron and positron solutions
carted hamiltonian is deduced acting in the
basis only from the electron states.
It is done to remove
so-called "zitterbewegung" [20].
\par b)
From above
mentioned hamiltonian
operator equation of the spin motion is deduced.
Then operators are replaced with the appropriate classical values,
i.e.  motion of a spin on classical trajectories is
considered.
 \par Under dispersion equation derivation it is not used any from
those assumptions and coincidence of the results seems
surprising.
\par Author is grateful to Dr. A.Ya. Silenko and
professor  V.V. Tikhomirov for the discussion of the
question.
\par \medskip

\centerline {\bf References}
\par
\noindent 1. V.V.  Tikhomirov  // Pis'ma Zh. Teor. Eksp. Fiz.
1995.  v.61.  p. 177; 690.
\par
\noindent
2. V.G.  Baryshevsky, V.L.
Lyboshitz  // Yad. Fiz. 1965. v.2. p.666.
\par \noindent 3.
C. Itzykson, J.-B. Zuber  Quantum field theory. v.1.  \par
McGraw-Hill Book company, 1978.
\par \noindent 4. L.D. Landau,
E.M. Lifshitz
Field theory. Moscow,1988.
 \par \noindent 5. V.B.
Berestetzky, E.M. Lifshitz,
L.P. Pitaevsky
\par Quantum
electrodynamics. Moscow,1989.
\par \noindent 6.
L.D. Landau,
E.M. Lifshitz
Mechanics. Moscow,1988.
 \par
\noindent 7. R.H. Good  // Phys. Rev.1962. v.125. N.6 p.2112-1115.
\par
\noindent 8. P. Nuborg  // Nuovo Cimento. 1964. v.31. p.1209-1228.
\par
\noindent 9. S.L. Cherkas  // Vesti Acad. Nauk Belarusi
\par ser. Fiz. 1994. N. 2. p. 70-75.
\par \noindent
10. A. Silenko   // Teor. Matem. Fiz. 1995. v.105 N.1 p. 46-54.
\par
\noindent 11. E. Plathe  // Suppl. Nuovo Cimento. 1966. v. $ 4. N1 $
p.246-275.  \par \noindent 12. E. Plathe  // Suppl. Nuovo Cimento. 1967. v.
$ 5. N3 $ p.944-953.  \par \noindent 13. K. Rafanelli  // Nuovo Cimento A.
1970. v. $ 67. N1 $ p.48-53.  \par \noindent 14. M. Lax  // Rev. Mod. Phys.
1951. v. $ 23 N4 $ p.287-310.  \par \noindent 15. V. G. Baryshevsky  Nuclear
optics of the polarized medium.  \par Moscow, 1995.  \par \noindent 16. V.G.
Baryshevsky  // Yad.Fiz. 1988. v.48. p. 1063-1066.
 \par
\noindent 17. I.M. Ternov {\it et al} // Zh. Teor. Eksp. Fiz.
1978 v. 74
p.  1201-1207.
\par \noindent 18. V.N. Baier  {\it et al} // Zh. Teor. Eksp, Fiz. 1975
v. p. 783-799.
\par \noindent 19. I.M. Ternov, V.R. Halilov
V.N. Rodionov
\par Interaction of charged particles with strong
electromagnetic field.  Moscow,1982, p.227-233.
 \par \noindent 20. I.M. Ternov
Introduction to the spin physics of relatiistic particles. Moscow,1997,
p.45.  \end {document}